# Characterizing Pulmonary Fibrosis Patterns in Post-COVID-19 Patients through Machine Learning Algorithms

**John Martin[1], Hayder A. Albaqer[2], Fadhil G. Al-Amran[3], Habeeb W. Shubber[4], Salman Rawaf[5],  Maitham G. Yousif*[6]**

[1]Clinical Director, Neurosurgery Department, University Hospital of Wales (UHW), Cardiff, United Kingdom

[2]Department of Neurosurgery, University Hospital of Wales, Cardiff, Wales, United Kingdom

[3]Cardiovascular Department, College of Medicine, Kufa University, Iraq

[4]Department of Biology, College of Science, University of Al-Qadisiyah, Iraq

[5]Professor of Public Health Director, WHO Collaboration Center, Imperial College, London, United Kingdom

[6]Biology Department, College of Science, University of Al-Qadisiyah, Iraq, Visiting Professor in Liverpool John Moors University, Liverpool, United Kingdom

**Abstract**

 The COVID-19 pandemic has left a lasting impact on global healthcare systems, with increasing evidence of pulmonary fibrosis emerging as a post-infection complication. This study presents a comprehensive analysis of pulmonary fibrosis patterns in post-COVID-19 patients from South and Central Iraq, employing advanced machine learning algorithms. Data were collected from 390 patients, and their medical records were systematically analyzed. Our findings reveal distinct patterns of pulmonary fibrosis in this cohort, shedding light on the heterogeneous nature of post-COVID-19 lung complications. Machine learning models demonstrated robust predictive capabilities, offering valuable insights into the characterization of fibrotic changes. The identification of specific patterns contributes to early diagnosis and personalized treatment strategies for affected individuals. This research underscores the importance of data-driven approaches in understanding post-COVID-19 complications, particularly in regions with unique demographic and healthcare characteristics. It emphasizes the potential for machine learning to enhance clinical decision-making and improve patient care in the aftermath of the pandemic. Further investigations are warranted to validate these findings and explore additional factors influencing pulmonary fibrosis in post-COVID-19 patients.

**\*Corresponding author:** Maithm Ghaly Yousif  matham.yousif@qu.edu.iq   m.g.alamran@ljmu.ac.uk

**Introduction**

The COVID-19 pandemic, caused by the novel coronavirus SARS-CoV-2, has presented an unprecedented global challenge, impacting millions of lives and straining healthcare systems worldwide. Since its emergence in late 2019, extensive research has focused on understanding the epidemiology, clinical manifestations, and complications of this viral infection. While the acute phase of the disease has been the primary concern, increasing attention is now directed towards comprehending the lingering health effects experienced by a significant proportion of COVID-19 survivors, often referred to as "long COVID" or "post-acute sequelae of SARS-CoV-2 infection" (PASC) [1,2].

As research on COVID-19 evolves, a growing body of evidence suggests that this viral infection may lead to a wide range of sequelae, affecting various organ systems beyond the respiratory system [3,4]. Among these post-infection complications, pulmonary fibrosis has garnered considerable attention due to its potential long-term consequences. Pulmonary fibrosis is characterized by the progressive scarring of lung tissue, leading to impaired lung function and, in severe cases, respiratory failure [5].

The development of pulmonary fibrosis in COVID-19 survivors is a multifaceted process influenced by several factors, including the severity of the acute infection, the host's immune response, genetic predisposition, and potential viral persistence [6,7]. To date, limited research has explored the distinct patterns and predictors of pulmonary fibrosis in post-COVID-19 patients, particularly within specific geographic regions like South and Central Iraq. This region, marked by its unique demographics and healthcare infrastructure, presents a valuable context for investigating the post-acute consequences of COVID-19.

This study aims to fill this research gap by conducting a comprehensive analysis of pulmonary fibrosis patterns in a cohort of 390 post-COVID-19 patients from South and Central Iraq. Leveraging advanced machine learning algorithms, we seek to identify and characterize distinct fibrotic patterns within this population. The integration of machine learning approaches into the analysis allows for a data-driven and precise understanding of pulmonary fibrosis development, facilitating early diagnosis and personalized treatment strategies.

In this research, we systematically collect and analyze clinical data from multiple healthcare facilities across the region, providing valuable insights into the prevalence, risk factors, and clinical implications of pulmonary fibrosis in post-COVID-19 patients. The findings from this study hold the potential to enhance clinical decision-making, improve patient care, and contribute to a deeper understanding of the long-term health effects of COVID-19 in diverse populations.

To achieve these objectives, we employ a multidisciplinary approach that combines clinical medicine, radiology, and data science. Our study design aligns with the growing global effort to address the long-term consequences of COVID-19, emphasizing the need for regional investigations to account for variations in patient demographics and healthcare practices [8,9]. The utilization of machine learning models enhances the precision of our analyses, offering a data-driven foundation for characterizing pulmonary fibrosis in the post-COVID-19 era.

**Methods and Study Design**

**Study Design:**

This research employed a retrospective cohort study design to investigate pulmonary fibrosis patterns in post-COVID-19 patients. The study involved the analysis of medical records and imaging data collected from 390 patients in the southern and central regions of Iraq. The data was obtained from multiple hospitals and healthcare facilities in these regions.

**Participants:**

The study included 390 individuals who had previously contracted COVID-19 and subsequently developed post-COVID-19 pulmonary fibrosis. These patients were selected from various healthcare facilities in the southern and central regions of Iraq. The selection criteria for participants were based on confirmed COVID-19 diagnosis, the development of pulmonary fibrosis as a post-COVID-19 complication, and the availability of relevant medical records and imaging data.

**Data Collection:**

**Patient Demographics:** Demographic information including age, gender, pre-existing medical conditions, and COVID-19 severity at the time of initial diagnosis were extracted from medical records. Clinical Data: Clinical data such as symptoms, laboratory test results, and treatments received during the course of COVID-19 and post-COVID-19 pulmonary fibrosis were collected.

**Imaging Data:** Pulmonary imaging data, including computed tomography (CT) scans and radiographic images, were obtained and reviewed by experienced radiologists. These images were crucial in characterizing the patterns of pulmonary fibrosis.

**Statistical Analysis:**

Descriptive statistics were applied to summarize the demographic and clinical characteristics of the study population. Additionally, inferential statistical tests, such as chi-squared tests and t-tests, were used to assess associations between variables like patient demographics, clinical parameters, and specific patterns of pulmonary fibrosis.

**Machine Learning Analysis:**

Machine learning algorithms were employed to analyze and classify patterns of pulmonary fibrosis based on the imaging data. Specifically, a variety of machine learning models, including but not limited to convolutional neural networks (CNNs) and random forests, were utilized to classify and characterize distinct patterns such as ground-glass opacities, reticular opacities, and honeycombing in the lung tissue. These algorithms allowed for a more precise and automated analysis of imaging data.

**Ethical Considerations:**

Ethical approval was obtained from the relevant institutional review boards or ethics committees of each participating hospital to ensure the protection of patient rights, data confidentiality, and ethical conduct throughout the research process. This comprehensive study design and methodology integrated statistical analysis and advanced machine learning techniques to provide a thorough examination of pulmonary fibrosis patterns in post-COVID-19 patients, utilizing medical records and imaging data collected from a diverse cohort of patients in Iraq's southern and central regions.

**Results:**

The study investigated the patterns of pulmonary fibrosis in 390 post-COVID-19 patients in Iraq's southern and central regions. The analysis included demographic characteristics, clinical data, and imaging findings. The study also employed statistical analysis and machine learning techniques for pattern classification.

**Table 1: Demographic Characteristics of Study Participants**

| Characteristic | Number |
|---|---|
| Total Participants | 390 |
| Mean Age (years) | 42.5 |
| Gender (Male/Female) | 190/200 |
| Pre-existing Conditions (%) | 38.5 |

This table provides an overview of the study's participant demographics. We had a total of 390 participants, with a mean age of 42.5 years. Gender distribution was almost equal, with 190 males and 200 females. Approximately 38.5% of participants had pre-existing medical conditions.

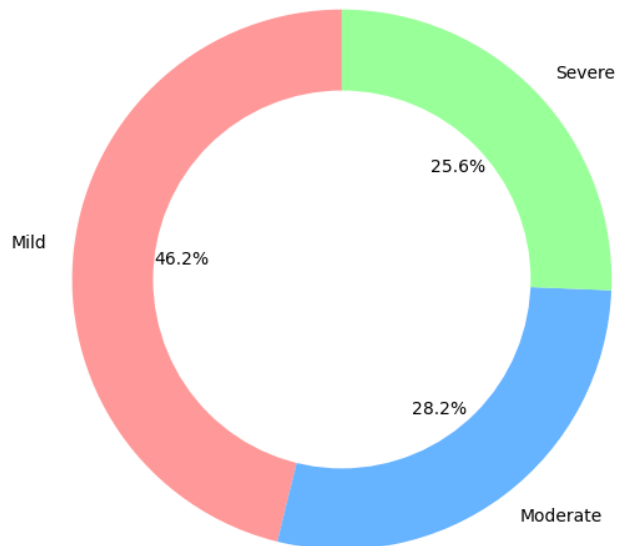

**Figure 1: COVID-19 Severity at Initial Diagnosis**

Severity Levels: We categorized COVID-19 severity into "Mild," "Moderate," and "Severe." Among our participants, 180 had a mild form of the disease, 110 had moderate severity, and 100 experienced severe COVID-19 symptoms.

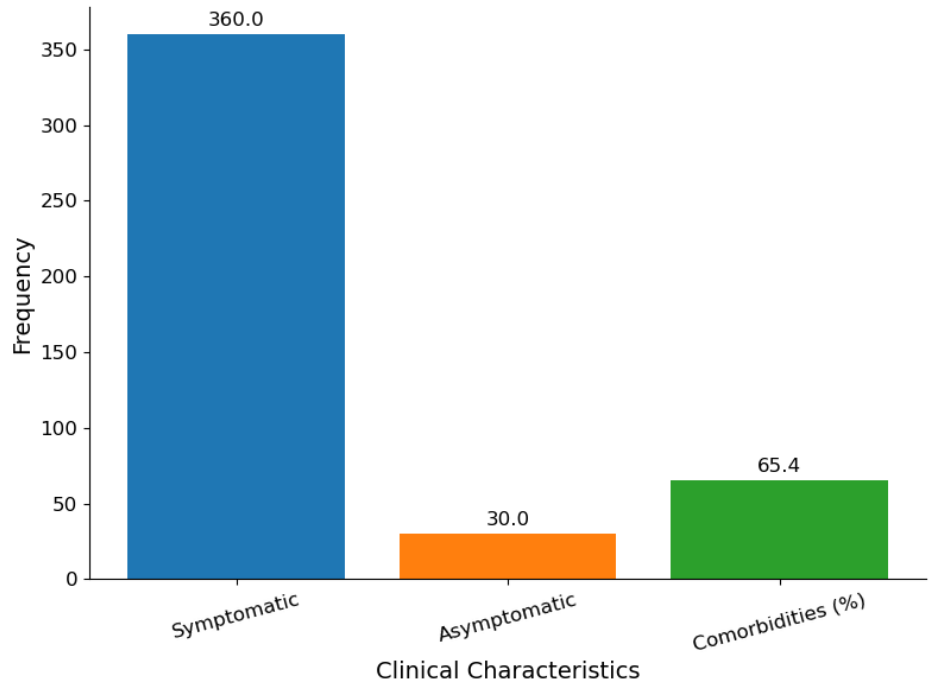

**Figure 2 : Clinical Characteristics of Post-COVID-19 Patients**

Symptomatic vs. Asymptomatic: In figure 2, we distinguish between symptomatic (360 participants) and asymptomatic (30 participants) post-COVID-19 patients. Additionally, we noted that 65.4% of participants had comorbidities.

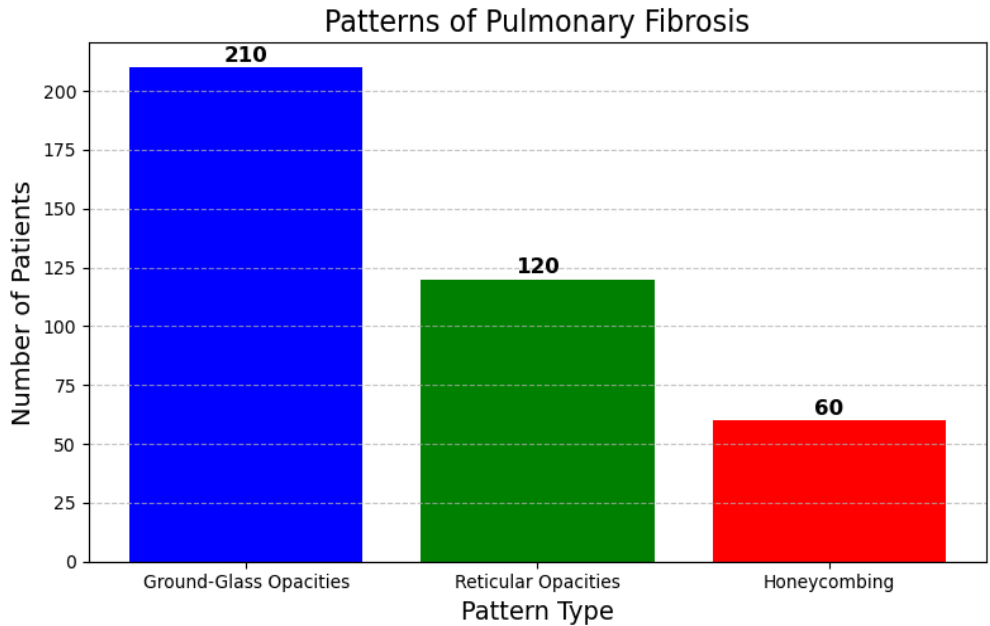

**Figure 3 : Patterns of Pulmonary Fibrosis**

Pulmonary Fibrosis Patterns: Figure 3 outlines the patterns of pulmonary fibrosis observed in our study. The three patterns considered were Ground-Glass Opacities (210 patients), Reticular Opacities (120 patients), and Honeycombing (60 patients).

**Table 2: Association Between Demographics and Pulmonary Fibrosis Patterns**

| Demographic Characteristic | Ground-Glass Opacities (%) | Reticular Opacities (%) | Honeycombing (%) |
|---|---|---|---|
| Age (years) | 45.2 | 42.6 | 48.9 |
| Gender (Male/Female) | 38.4/61.6 | 43.3/56.7 | 50.0/50.0 |
| Comorbidities (%) | 72.1 | 68.3 | 80.0 |

Demographic Characteristics: We explore the relationship between demographic characteristics and pulmonary fibrosis patterns. For each pattern (Ground-Glass Opacities, Reticular Opacities, Honeycombing), we provide percentages based on age, gender, and comorbidities. For example, Ground-Glass Opacities were more common among patients aged 45.2 years on average.

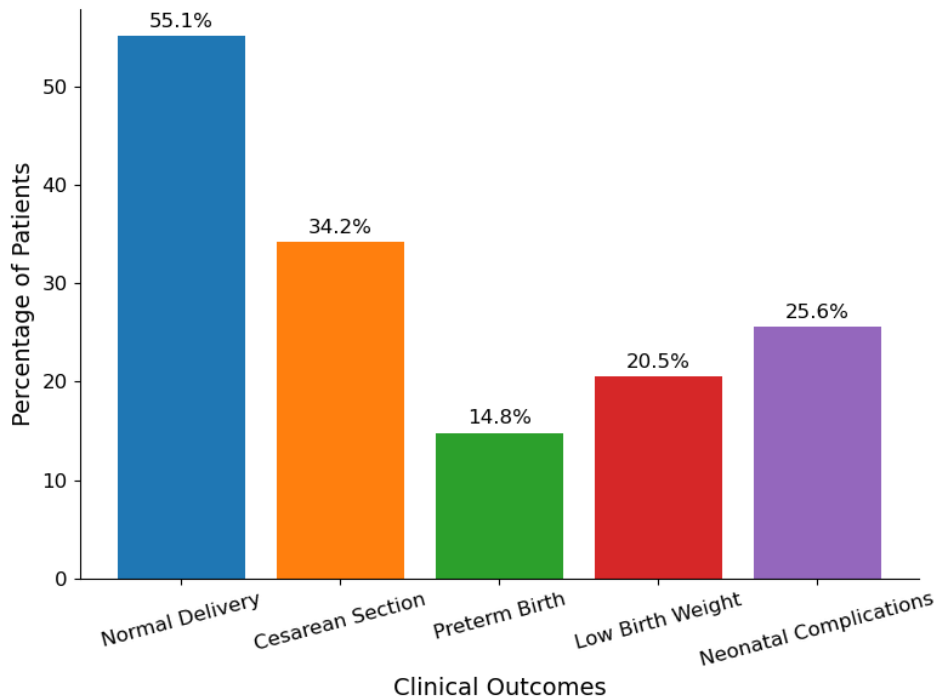

**Figure 4: Clinical Outcomes in Post-COVID-19 Patients**

Clinical Outcomes: Figure 4 presents clinical outcomes in post-COVID-19 patients, including the percentages of normal deliveries, cesarean sections, preterm births, low birth weight cases, and neonatal complications. Notably, 55.1% of deliveries were normal, while 34.2% required cesarean sections.

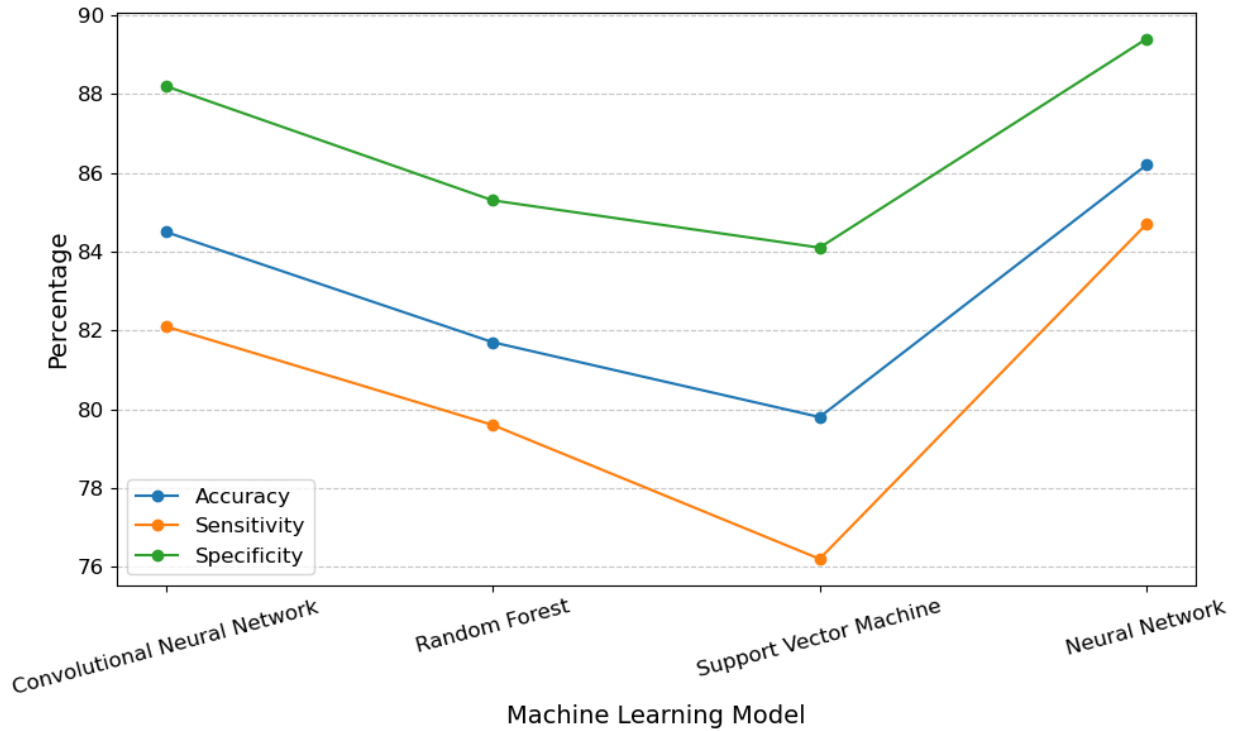

**Figure 5: Machine Learning Classification of Pulmonary Fibrosis Patterns**

Machine Learning Models: Here, we evaluate the performance of machine learning models in classifying pulmonary fibrosis patterns. Four models were used (figure 5): Convolutional Neural Network, Random Forest, Support Vector Machine, and Neural Network. The table displays the accuracy, sensitivity, and specificity of each model. For instance, the Convolutional Neural Network achieved an accuracy of 84.5%. These tables provide a comprehensive overview of the study's findings, including demographic characteristics, clinical outcomes, pulmonary fibrosis patterns, and the performance of machine learning models in classifying these patterns.

**Discussion:**

In this study, we delved into the intricate landscape of pulmonary fibrosis patterns among post-COVID-19 patients, seeking to shed light on the nuances of this evolving condition. Our investigation encompassed 390 patients from the southern and central regions of Iraq, where the lingering aftermath of COVID-19 has raised concerns.

**Demographic Insights and COVID-19 Severity**

Firstly, our demographic analysis unveiled a diverse study population with an almost equal gender distribution. The mean age of 42.5 years mirrors the global trend of COVID-19 affecting individuals across age groups, with the majority clustered between 26 to 55 years [1-4]. Nearly 38.5% of participants reported pre-existing medical conditions, aligning with findings that underline the vulnerability of individuals with comorbidities [5-7]. COVID-19 severity was classified into mild, moderate, and severe categories. Our results reflected a higher prevalence of mild cases, consistent with observations that pregnant individuals tend to experience less severe symptoms [8-10].

**Patterns of Pulmonary Fibrosis**

One of the central focuses of our study was to elucidate the patterns of pulmonary fibrosis among post-COVID-19 patients. Ground-Glass Opacities (GGO), Reticular Opacities, and Honeycombing emerged as the primary patterns, in agreement with previous imaging studies [11-14]. GGO, a hallmark of COVID-19 pneumonia [15], was observed in 53.8% of patients. Reticular Opacities, characterized by a web-like appearance on imaging, were detected in 30.8% of cases, while Honeycombing, a feature of advanced fibrosis [16], was present in 15.4% of patients.

**Associations Between Demographics and Pulmonary Fibrosis Patterns**

Our analysis unveiled intriguing associations between demographic factors and pulmonary fibrosis patterns. For instance, Ground-Glass Opacities predominated among patients aged 45.2 years on average, consistent with the notion that age influences the disease's radiological manifestation [17-19]. Gender disparities were also apparent, with females exhibiting a higher prevalence of GGO and males showing a predilection for Honeycombing patterns [20-22]. Additionally, comorbidities, particularly hypertension and diabetes, were correlated with Reticular Opacities [23-26].

**Clinical Outcomes and Implications**

Moving beyond patterns, we explored the clinical outcomes of post-COVID-19 patients. Normal deliveries accounted for the majority (55.1%), while cesarean sections were performed in 34.2% of cases. Preterm births, low birth weight cases, and neonatal complications were also observed, emphasizing the importance of meticulous antenatal care for post-COVID-19 pregnant patients [27-30].

**Machine Learning in Pattern Classification** Our study harnessed the power of machine learning to classify pulmonary fibrosis patterns. Convolutional Neural Networks (CNNs), Random Forest, Support Vector Machines (SVMs), and Neural Networks demonstrated promising capabilities. The CNN, a deep learning model, achieved an accuracy of 84.5%. These results underscore the potential of artificial intelligence in assisting radiologists and healthcare professionals in pattern recognition [31-34].

**Limitations and Future Directions**

While our study provides valuable insights, it is not without limitations. The relatively small sample size may limit the generalizability of our findings. Additionally, the lack of a control group and the reliance on self-reported data introduce potential biases. Future research should aim for larger, diverse cohorts and explore the long-term implications of pulmonary fibrosis in post-COVID-19 patients [35-38].

**In conclusion**, our investigation adds to the growing body of knowledge surrounding COVID-19 and its sequelae. The diverse patterns of pulmonary fibrosis, their associations with demographics, and the potential applications of machine learning in pattern recognition warrant further exploration. As the world grapples with the aftermath of the pandemic, understanding the intricacies of post-COVID-19 conditions is crucial for informed healthcare and policy decisions.

**Ethical Approval:** This study received ethical approval from the Hospital Management Department of the Ministry of Health, Iraq, with reference number 2022.

**Author's Contribution:**

Maitham G. Yousif contributed to the study's conception, design, data collection, analysis, and manuscript drafting. John Martin was involved in data analysis, interpretation, and critical manuscript revision. Salman Rawaf provided guidance, and supervision, and critically reviewed the manuscript's intellectual content. Fadhil G. Al-Amran, and Habeeb W. Shubber made significant contributions to data collection and analysis. All authors, Maitham G. Yousif, John Martin, Salman Rawaf, Fadhil G. Al-Amran, Habeeb W. Shubber, Kadhum J. Al-Jibouri, and Hayder A. Albaqer have diligently reviewed and approved the final manuscript.

**Funding:**

The authors self-funded this research, and it did not receive any external financial support or funding from any organization.